\documentclass[prd,unsortedaddress,superscriptaddress,showpacs,a4paper,
nofootinbib]{revtex4}
\usepackage{epsfig}
\def\beq{\begin{equation}}
\def\enq{\end{equation}}
\def\beqa{\begin{eqnarray}}
\def\enqa{\end{eqnarray}}
\def\nnb{\nonumber}

\def\MeV{\nobreak\,\mbox{MeV}}
\def\GeV{\nobreak\,\mbox{GeV}}

\def\qq{\lag\bar{q}q\rag}
\def\uu{\lag\bar{u}u\rag}
\def\dd{\lag\bar{d}d\rag}

\def\mix{\lag\bar{q}g\si.Gq\rag}

\def\Gd{\lag g^2G^2\rag}
\def\G3{\lag g^3G^3\rag}

\def\ga{\gamma}

\def\rh{\rho}
\def\si{\sigma}

\def\al{\alpha}
\def\be{\beta}
\def\alma{\alpha_{max}}
\def\almi{\alpha_{min}}
\def\bemi{\beta_{min}}
\def\lb{\label}
\def\nn{\nonumber}
\def\kab{\left[(\al+\be)m_c^2-\al\be s\right]}
\newcommand{\rag}{\rangle}
\newcommand{\lag}{\langle}

\begin{document}

\title{\sc 
Can the $X(3872)$ be a $1^{++}$ four-quark state?
}
\author{R.D.~Matheus}
\email{matheus@if.usp.br}
\affiliation{Instituto de F\'{\i}sica, Universidade de S\~{a}o Paulo, 
C.P. 66318, 05389-970 S\~{a}o Paulo, SP, Brazil}
\author{S.~Narison}
\email{snarison@yahoo.fr}
\affiliation{ Laboratoire de Physique Th\'{e}orique et Astrophysiques,
CNRS-IN2P3-UM2, Place Eug\`ene Bataillon,
34095 - Montpellier Cedex 05, France}
\author{M.~Nielsen}
\email{mnielsen@if.usp.br}
\affiliation{Instituto de F\'{\i}sica, Universidade de S\~{a}o Paulo, 
C.P. 66318, 05389-970 S\~{a}o Paulo, SP, Brazil}
\author{J.-M.~Richard}
\email{jean-marc.richard@lpsc.in2p3.fr}
\affiliation{Laboratoire de Physique Subatomique
 et Cosmologie, Universit\'e Joseph 
Fourier--IN2P3-CNRS,
53, avenue des Martyrs, 38026 Grenoble cedex, France}

\begin{abstract}
We use QCD spectral sum rules to test the nature of the meson
$X(3872)$, assumed to be an exotic four-quark $(c\bar{c}q\bar{q})$
state with $J^{PC}=1^{++}$. For definiteness, we work with the current 
proposed recently by Maiani et al \cite{Maiani},
at leading order in $\alpha_s$, consider the contributions of higher
dimension condensates and keep terms which are linear in
the light quark mass $m_q$.  
We find $M_X=(3925\pm 127)$  MeV which is
compatible, within the errors, with the experimental candidate X(3872), 
while  the
SU(3) breaking-terms lead to an unusual mass-splitting $M_{X^{s}}-M_X=-
(61\pm 30)$ MeV.
The mass-difference between the neutral states due to isospin
violation of about $(2.6 \sim 3.9)$ MeV is  smaller than the value $(8\pm 3)$ MeV 
proposed in \cite{Maiani}.  
For the $b$-quark, we predict 
$M_{X_b}= (10144\pm 106)~{\rm MeV}$ for the $X_b(b\bar{b}q \bar{q})$,  which
is much below the $\bar BB^*$ threshold in contrast to the $\bar BB^*$
molecule  prediction \cite{Swanson}, and for the $X_b^s(b\bar{b}s \bar{s})$, a 
mass-splitting $M_{X^s_{b}}-M_{X_b}=-(121\pm 182)$ MeV.  Our analysis also 
indicates 
that the mass-splitting between the ground state and the radial excitation 
of about $(225\sim 250$) MeV is much smaller than in the case of ordinary mesons
and is (within the errors) flavour-independent.
We also extract the decay
constants, analogous to $f_\pi$, of such mesons,
which are useful for further studies of their  leptonic and hadronic decay 
widths. The uncertainties of our estimates are mainly
due to the ones from the $c$ and $b$ quark masses. 
\end{abstract}

\maketitle

\section{Introduction}
In August 2003, Belle reported evidence for a new narrow state in the
decay $B^+\!\rightarrow\!X(3872)K^+\rightarrow\!J/\psi\pi^+\pi^- K^+$
\cite{BELLE}, which has been confirmed by three other experiments
\cite{Xexpts}.  The $X(3872)$ is the best studied of the new
$c\bar{c}$-associated states, $X(3872)$, $X(3940)$, $Y(4260)$, etc.
\cite{Swanson}.  It has a mass of 3872 MeV and a very narrow width
$\Gamma < 2.3$ MeV at 95\%.  Upon discovery, $X(3872)$ seemed a likely
candidate for $\psi_{2}(^3D_2)$ or $\psi_{3}(^3D_3)$ \cite{choi}, but
the expected radiative transitions to $\chi_{c}$ states have never
been seen.  The $\pi\pi$ mass spectrum favors high dipion masses,
suggesting a $J/\psi\,\rho$ decay that is incompatible with the
identification of $X(3872) \to \pi^{+}\pi^{-}\,J/\psi$ as the strong
decay of a pure isoscalar state.  Belle's observation of the decay
$X(3872) \to J/\psi\,\gamma$ \cite{belleE} determines $C=+$, opposite
to the charge-conjugation of the leading charmonium candidates.  The
same paper \cite{belleE} also reports the observation of the $X$
decaying to $J/\psi\,\pi^+\pi^-\pi^0$, with a rate which is comparable
to that of the $J/\psi\pi^+\pi^-$ mode.  This decay suggests an
appreciable transition rate to $J/\psi\,\omega$ and establishes
sizeable isospin violating effects.  Finally, an analysis of angular
distributions supports the assignment $J^{PC} = 1^{++}$, but the mass
of $X(3872)$ is too low to be gracefully identified with the $2~^3P_1$
charmonium state. More recently, the Belle collaboration reported a peak in
$D^0\bar{D}{}^0\pi^0$ which can be interpreted as the dominant decay
mode of the $X$ \protect\cite{Gokhroo:2006bt}.\@

The anomalous nature of the $X$ has led to many speculations:
tetraquark \cite{Maiani,hrs}, cusp \cite{bugg}, hybrid \cite{li}, or
glueball \cite{seth}.  Another explanation is that the $X(3872)$ is a
$D\bar D^*$ bound state \cite{tor,clopa,wong,pasu,ess}, as predicted
before its discovery.

In this work we use QCD spectral sum rules (QSSR) (the Borel/Laplace Sum
Rules (LSR) \cite{svz,rry,SNB} and Finite Energy Sum Rules (FESR)
\cite{SNB,RAF,KRAS}), to study the two-point functions of the axial
vector meson, $X(3872)$, assumed to be a four-quark state.  In
previous calculations, the Sum Rule (SR) approach was used to study
the light scalar mesons \cite{LATORRE,SN4,sca,koch} and the
$D_{sJ}^+(2317)$ meson \cite{pec,OTHERA}, considered as four-quark states and
a good agreement with the experimental masses was obtained.  However,
the tests were not decisive as the usual quark--antiquark assignments
also provide predictions consistent with data and more importantly
with chiral symmetry expectations \cite{SN4,SNHEAVY,SNB,OTHER}.  In the
four-quark scenario, scalar mesons can be considered as S-wave bound
states of diquark-antidiquark pairs, where the diquark was taken to be
a spin zero color anti-triplet.  Here we follow ref.~\cite{Maiani},
and consider the $X(3872)$ as the $J^{PC}=1^{++}$ state with the
symmetric spin distribution: $[cq]_{S=1}[\bar{c}\bar{q}]_{S=0}+
[cq]_{S=0}[\bar{c}\bar{q}]_{S=1}$.  Therefore, the corresponding
lowest-dimension interpolating operator for describing $X_q$ is given
by: 
\beq
j_\mu={i\epsilon_{abc}\epsilon_{dec}\over\sqrt{2}}[(q_a^TC\gamma_5c_b)
(\bar{q}_d\gamma_\mu C\bar{c}_e^T)+(q_a^TC\gamma_\mu c_b)
(\bar{q}_d\gamma_5C\bar{c}_e^T)]\;,
\label{field}
\enq
where $a,~b,~c,~...$ are color indices, $C$ is the charge conjugation
matrix and $q$ denotes a $u$ or $d$ quark.

In general, one should consider all possible combinations of 
different $1^{++}$ four-quark operators, similar to e.g. done in 
\cite{chinois}  
for the $0^{++}$ light mesons and consider their mixing under 
renormalizations \cite{TARRACH} from which one can form  renormalization 
group invariant 
(RGI) physical currents. However, we might expect that, working with a
particular choice of current given above will provide a general
feature of the four-quark model predictions for the $X(3872)$, provided that
we can work with quantities less affected by radiative corrections and where the
OPE converges quite well \footnote{In the well-known case of baryon sum rules,
a simplest choice of operator \cite{io1} and a more general choice 
\cite{DOSCH} 
have been given in the literature. Though technically apparently different, 
mainly for 
the region of convergence of the OPE, the two choices of interpolating currents
have provided the same predictions for the proton mass and mixed condensate
but only differs for values of higher dimension four-quark condensates.}
As pointed out in \cite{Maiani}, isospin forbidden decays are possible
if $X$ is not a pure isospin state.  Pure isospin states are: \beq
X(I=0)={X_u+X_d\over\sqrt{2}},\;\;\;\mbox{and}\;\;\;\;X(I=1)={X_u-X_d\over
\sqrt{2}}.  \enq If the physical states are just the mass eigenstates
$X_u$ or $X_d$, maximal isospin violations are possible.  Deviations
from these two ideal situations are described by a mixing angle
between $X_u$ and $X_d$ \cite{Maiani}: \beqa
X_l&=&X_u\cos{\theta}+X_d\sin{\theta}, \nonumber\\
X_h&=&-X_u\sin{\theta}+X_d\cos{\theta}.
\label{lh}
\enqa

In ref.~\cite{Maiani}, by considering the $X$ decays into two and
three pions, a mixing angle $\theta\sim20^\circ$ is deduced and a mass
difference 
\beq m(X_h)-m(X_l)=(8\pm3)\MeV.
\label{difma}
\enq
In this work, we want to test in which conditions the results of the
sum rules are compatible with the above predictions.

\section{The QCD expression of the two-point correlator}
The SR are constructed from the two-point correlation function
\beq
\Pi_{\mu\nu}(q)=i\int d^4x ~e^{iq.x}\lag 0
|T[j_\mu(x)j^\dagger_\nu(0)]
|0\rag=-\Pi_1(q^2)(g_{\mu\nu}-{q_\mu q_\nu\over q^2})+\Pi_0(q^2){q_\mu
q_\nu\over q^2}.
\lb{2po}
\enq
Since the axial vector current is not conserved, the two functions,
$\Pi_1$ and $\Pi_0$, appearing in Eq.~(\ref{2po}) are independent and
have respectively the quantum numbers of the spin 1 and 0 mesons.

The fundamental assumption of the sum rules approach is the principle
of duality.  Specifically, we assume that there is an interval over
which the correlation function may be equivalently described at both
the quark and the hadron levels.  Therefore, on the one hand, we
calculate the correlation function at the quark level in terms of
quark and gluon fields.  On the other hand, the correlation function
is calculated at the hadronic level introducing hadron characteristics
such as masses and coupling constants.  At the quark level, the
complex structure of the QCD vacuum leads us to employ the Wilson's
operator product expansion (OPE).  The calculation of the
phenomenological side proceeds by inserting intermediate states for
the meson $X$.  Parametrizing the coupling of the axial vector meson
$1^{++}$, $X$, to the current, $j_\mu$, in Eq.~(\ref{field}) in terms
of the meson decay constant $f_X$ as:
\beq\label{eq: decay}
\lag 0 |
j_\mu|X\rag =\sqrt{2}f_XM_X^4\epsilon_\mu~, 
\enq
the phenomenological side
of Eq.~(\ref{2po}) can be written as 
\beq
\Pi_{\mu\nu}^{phen}(q^2)={2f_X^2M_X^8\over
M_X^2-q^2}\left(-g_{\mu\nu}+ {q_\mu q_\nu\over M_X^2}\right)
+\cdots\;, \lb{phe} \enq
where the Lorentz structure projects out the $1^{++}$ state.  The dots
denote higher axial-vector resonance contributions that will be
parametrized, as usual, through the introduction of a continuum
threshold parameter $s_0$.

In the OPE side, we work at leading order in $\alpha_s$ and consider the
contributions of condensates up to dimension eight.  We keep the term
which is linear in the light-quark mass $m_q$, in order to estimate
the mass difference in Eq.~(\ref{difma}).  Keeping the charm-quark
mass finite, we use the momentum-space expression for the charm-quark
propagator.  The light-quark part of the correlation function is
calculated in the coordinate-space, and then Fourier transformed to
the momentum space in $D$ dimensions.  The resulting light-quark part
is combined with the charm-quark part before it is dimensionally
regularized at $D=4$.

The correlation function, $\Pi_1$, in the OPE side can be written as a
dispersion relation:
\beq
\Pi_1^{OPE}(q^2)=\int_{4m_c^2}^\infty ds {\rho(s)\over s-q^2}\;,
\lb{ope}
\enq
where the spectral density is given by the imaginary part of the
correlation function: $\pi \rho(s)=\mbox{Im}[\Pi_1^{OPE}(s)]$.  After
making an inverse-Laplace (or Borel) transform of both sides, and
transferring the continuum contribution to the OPE side, the sum rule
for the axial vector meson $X$ 
up to dimension-eight condensates can
be written as \footnote{
We have not included the effects of a dimension 
2 term induced by the UV renormalon, \cite{ZAKA,ZAKA2},
which we expect to be numerically negligible like in the other channels 
\cite{SN2}, though this result needs to be checked. Instanton-like 
contributions which appear as a high-dimension
operators will also be neglected like some other higher dimension condensate 
effects.} : 
\beq 2f_X^2M_X^8e^{-M_X^2/M^2}=\int_{4m_c^2}^{s_0}ds~
e^{-s/M^2}~\rho(s)\; +\Pi_1^{mix\qq}(M^2)\;, \lb{sr} \enq
where 
\beq
\rho(s)=\rho^{pert}(s)+\rh^{m_q}(s)+\rh^{\qq}(s)+\rh^{\lag G^2\rag}
(s)+\rh^{mix}(s)+\rh^{\qq^2}(s)\;,
\lb{rhoeq}
\enq
with
\beqa\label{eq:pert}
&&\rho^{pert}(s)={1\over 2^{10} \pi^6}\int\limits_{\almi}^{\alma}
{d\al\over\alpha^3}
\int\limits_{\bemi}^{1-\al}{d\be\over\be^3}(1-\al-\be)(1+\al+\be)
\left[(\al+\be)
m_c^2-\al\be s\right]^4,
\nn\\
&&\rho^{m_q}(s)=-{m_q \over 2^3 \pi^4} \int\limits_{\almi}^{\alma}
{d\al\over\al} 
\bigg\{ -{\qq\over 2^2}{[m_{c}^2-\al(1-\al)s]^2 \over (1-\al)}
+ \int\limits_{\bemi}^{1-\al}{d\be\over\be}\kab \bigg[ 
- m_{c}^{2} \qq\nn\\
&&\qquad\qquad {}+{\qq\over 2^2} \kab
\nn+{m_c\over 2^5 \pi^2 \al \be^2}(3+\al+\be)(1-\al-\be)\kab^2 
\bigg] \bigg\},\nn\\
&&\rho^{\qq}(s)=-{m_c\qq\over 2^{5}\pi^4}\int\limits_{\almi}^{\alma}
{d\al\over\al^2}
\int\limits_{\bemi}^{1-\al}{d\be\over\be}(1+\al+\be)\left[(\al+\be)m_c^2-
\al\be s\right]^2,\nn\\
&&\rho^{\lag
G^2\rag}(s)={\Gd\over2^{9}3\pi^6}\int\limits_{\almi}^{\alma} d\al\!\!
\int\limits_{\bemi}^{1-\al}{d\be\over\be^2}\left[(\al+\be)m_c^2-\al\be
s\right]
\left[{m_c^2(1-(\al+\be)^2)\over\be}-
{(1-2\al-2\be)\over2\al}\left[(\al+\be)m_c^2-\al\be s\right]
\right]. \nnb\\
\enqa
where the integration limits are given by $\almi=({1-\sqrt{1-
4m_c^2/s})/2}$, $\alma=({1+\sqrt{1-4m_c^2/s})/2}$ and $(\bemi={\al
m_c^2)/( s\al-m_c^2)}$.
We have also included the dominant contributions from the dimension-five
condensates:
\beqa
\rho^{mix}(s)={m_c\mix\over 2^{6}\pi^4}\int\limits_{\almi}^{\alma}
d\al
\bigg[-{2\over\al}(m_c^2-\al(1-\al)s)
+\int\limits_{\bemi}^{1-\al}d\be\left[(\al+\be)m_c^2-\al\be
s\right]\left({1\over
\al}+{\al+\be\over\be^2}\right)\bigg],
\enqa
where the contribution of dimension-six condensates $\lag g^3 G^3\rag$ is
neglected, since assumed to be suppressed by   the loop factor $1/16\pi^2$. 
The usual estimate $\lag g^3 G^3\rag\simeq 1$GeV$^2\lag \alpha_s G^2\rag$ 
\cite{SNB} would deserve to be checked in more detail.
We have included the contribution of the dimension-six four-quark condensate:
\beqa
\rho^{\qq^2}(s)={m_c^2\qq^2\over 12\pi^2}\sqrt{s-4m_c^2\over
s},
\enqa
and (for completeness) a part of the dimension-8 condensate contributions 
\footnote{We should note that a complete evaluation of these contributions 
require more involved analysis including a non-trivial choice
of the factorization assumption basis \cite{BAGAN}. We wish that we can 
perform this analysis in the future.}:
\beqa
\Pi_1^{mix\qq}(M^2)=-{m_c^2\mix\qq\over 24\pi^2}\int_0^1
d\al\,\bigg[1+{m_c^2
\over\al(1-\al) M^2}-{1\over2(1-\al)}\bigg]\,\exp\!\left[{-{m_c^2
\over\al(1-\al)M^2}}\right].
\label{dim8}
\enqa
\boldmath\section{LSR  predictions of  $M_X$} \unboldmath
\begin{figure}[h] 
\centerline{\epsfig{figure=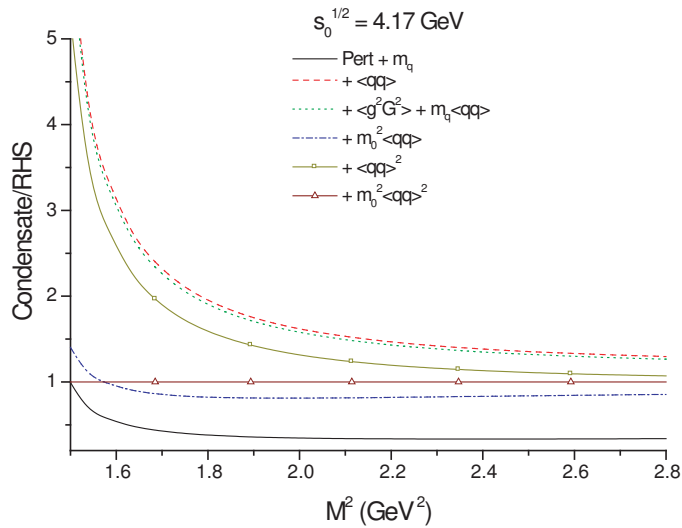,height=70mm}}
\caption{The OPE convergence in the region $1.6 \leq M^2 \leq
2.8~\GeV^2$ for $s_0^{1/2} = 4.17$ GeV. We start with the perturbative
contribution (plus a very small $m_q$ contribution) and each
subsequent line represents the addition of one extra condensate
dimension in the expansion.}
\label{figconvmar} 
\end{figure} 

In order to extract the mass $M_X$ without worrying about the value of
the decay constant $f_X$, we take the derivative of Eq.~(\ref{sr})
with respect to $1/M^2$, divide the result by Eq.~(\ref{sr}) and
obtain:
\beq
M_X^2={\int_{4 m_c^2}^{s_0}ds ~e^{-s/M^2}~s~\rho(s)\over\int_{4
m_c^2}^{s_0}
ds ~e^{-s/M^2}~\rho(s)}\;.
\lb{m2}
\enq
This quantity has the advantage to be less sensitive to the perturbative 
radiative corrections than the individual
moments. Therefore, we expect that our results obtained  to leading order 
in $\alpha_s$ will be quite accurate.

In the numerical analysis of the sum rules, the values used for the
quark
masses and condensates are (see e.g.  \cite{SNB,SNCB,narpdg,SNG}) 
\footnote{To leading order approximation in $\alpha_s$, at which we 
are working, we do not consistently consider the running scale dependence of these 
parameters. We shall use here the values of the quark masses obtained within the same
QCD spectral sum rules methods compiled in \cite{SNB}.  They are defined in
the ${\overline{MS}}$-scheme, and have obtained   
within the same truncation of the QCD series from different channels and by different authors. }
: 
\beqa\label{qcdparam}
&&m_c(m_c)=(1.23\pm 0.05)\,\GeV ,~~~~~ ~~m_b(m_b)=(4.24\pm 0.06)\,\GeV ,\nnb\\
&&m_u=2.3\,\MeV,~~~~~~~~~~~~~~~~~~~~~~~~ ~~m_d=6.4\,\MeV, \nnb\\
&&m_q\equiv {(m_u + m_d)/2}=4.3\,\MeV,~~~~\lag\bar{q}q\rag=\,-(0.23\pm0.03)^3
\,\GeV^3,
\nnb\\
&&m_s= 100~\MeV,~~~~~~~~~~~~~~~~~~~~~~~~\lag\bar{s}s\rag/\lag\bar{q}q\rag=
0.8\pm0.2,
\nnb\\
&&\lag\bar{q}g\si.Gq\rag=m_0^2\lag\bar{q}q\rag ~~~~~~{\rm with}~~~~~~~~~m_0^2=
0.8\,\GeV^2,\nnb\\
&&\lag g^2G^2\rag=0.88~\GeV^4.
\enqa
We evaluate the sum rules in the range $2.0 \leq M^2 \leq 2.8$ for
two values of $s_0$: $s_0^{1/2} = 4.1$ GeV, $s_0^{1/2} = 4.2$ GeV.

\begin{figure}[h] 
\centerline{\epsfig{figure=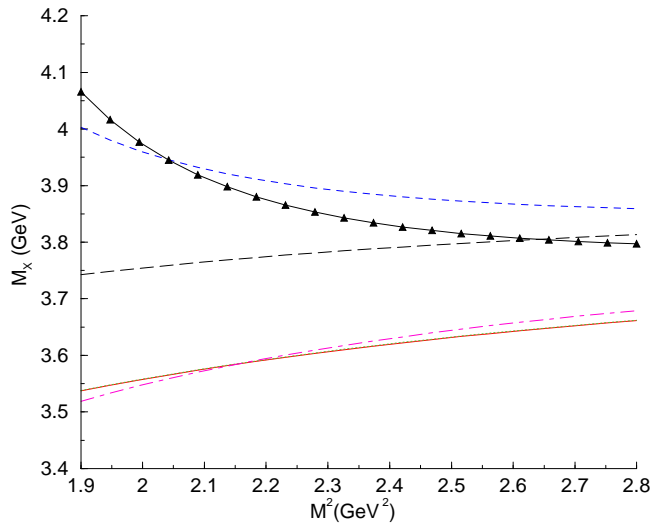,height=70mm}}
\caption{The OPE convergence for $M_X$ in the region $1.9 \leq M^2 \leq
2.8~\GeV^2$ for $s_0^{1/2} = 4.1$ GeV. We start with the perturbative
contribution plus a very small $m_q$ contribution (long-dashed line) and each
subsequent line represents the addition of one extra condensate
dimension in the expansion: $+\langle \bar{q}q\rangle$ (solid line),
$+\langle g^2G^2\rangle+ m_q\langle \bar{q}q\rangle$ (dotted-line in top of 
the solid line),
$+ \langle \bar{q}q\rangle^2$ (dashed line), $+ m_0^2\langle \bar{q}q
\rangle$ (solid line with triangles).}
\label{figconvmx} 
\end{figure}

\begin{figure}[h] 
\centerline{\epsfig{figure=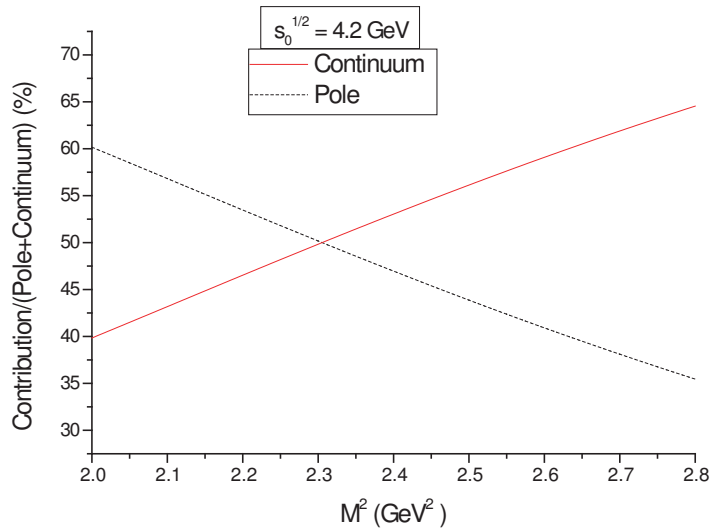,height=70mm}}
\caption{The dashed line shows the relative pole contribution (the
pole contribution divided by the total, pole plus continuum,
contribution) and the solid line shows the relative continuum
contribution.}
\label{figpvc} 
\end{figure}

\begin{figure}[h] 
\centerline{\epsfig{figure=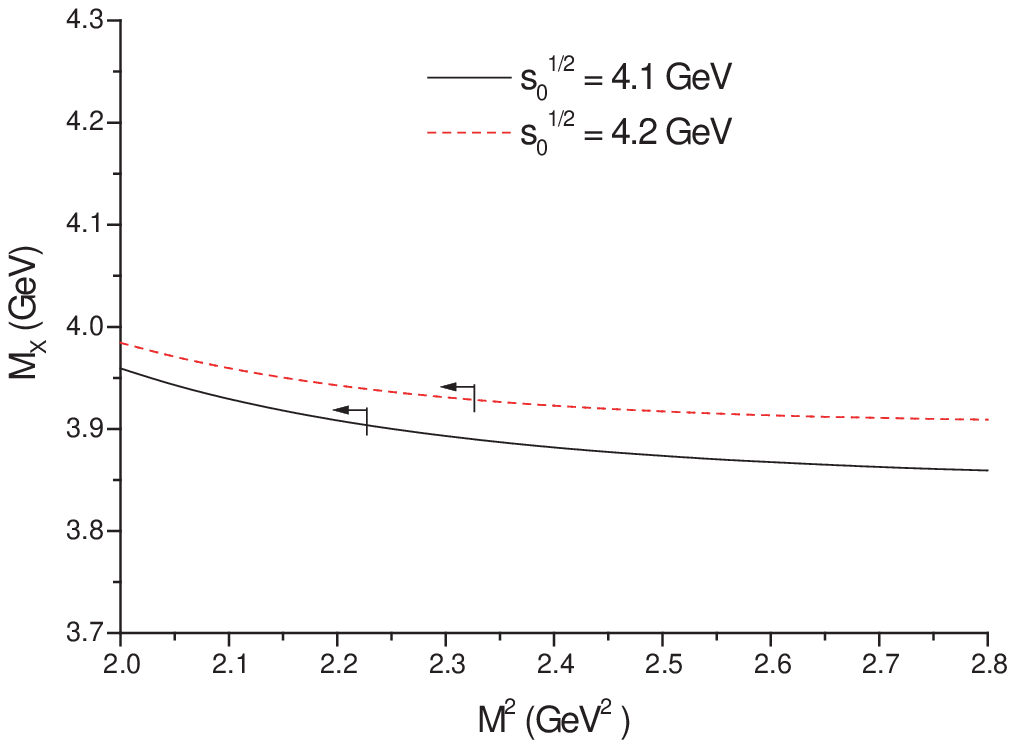,height=70mm}}
\caption{The $X$ meson mass as a function of the sum rule parameter
($M^2$) for different values of the continuum threshold: $s_0^{1/2} =
4.1$ GeV (solid line) and $s_0^{1/2} = 4.2$ GeV (dashed line).  The 
arrows indicate the region allowed for the sum rules: the lower limit 
(cut below 2.0 GeV$^2$) is given by OPE convergence requirement and the 
upper limit by the dominance of the QCD pole contribution.}
\label{figmx} 
\end{figure}

Comparing the relative contribution of each term in Eqs.~(\ref{eq:pert}) to 
(\ref{dim8}),
to the right hand side of Eq.~(\ref{sr}) we obtain a quite good OPE
convergence for $M^2 > 1.9$ GeV$^2$, as can be seen in
Fig.~\ref{figconvmar}.  This analysis allows us to determine the lower
limit constraint for $M^2$ in the sum rules window.  This figure also
shows that, although there is a change of sign between
dimension-six and dimension-eight condensates contributions, the
contribution of the latter being smaller, where, we have assumed, in 
Fig.~\ref{figconvmar} to  Fig.~\ref{figmx}, the validity of
the vacuum saturation for these condensates. The relatively small contribution 
of the dimension-eight condensates may justify  the validity
of our approximation, unlike in the case of the 5-quark current
correlator, as noticed in \cite{oganes}. However, the partial compensation of 
these two terms indicate the sensitivity of the central value of the mass 
prediction on the way the OPE is truncated.

In Fig.~\ref{figconvmx} are shown the contributions of the individual 
condensates  to $M_X$ obtained from Eq.~(\ref{m2}), 
From Fig.~\ref{figconvmx}, it appears that the results oscillates around
the perturbative result, and that the results obtained up to dimension-5 are 
very close to the ones obtained up to dimension-8. For definiteness,  the 
value of 
$M_X$ obtained by including the dimension-5 mixed condensate will be considered
as the final  prediction from the LSR, and
the effects of the higher condensates as the error due to the truncation of 
the OPE.

We get an upper limit constraint for $M^2$ by imposing the rigorous constraint 
that the QCD continuum contribution should be smaller than the pole 
contribution\footnote{More restrictive conditions 
are sometimes imposed in the literature, where, for example, it is required that 
the continuum contribution is smaller than 30 \% of the total contribution.
In this case no sum rule-window is allowed.  In our analysis, we use a 
less restrictive criterion, 
having in mind that the role of the continuum is expected to be larger for 
high-dimensional current operators than in the 
usual $\rho$-meson channel, as indicated by different sum rules 
analyses in the existing literature.}.
The maximum value of $M^2$ for which this constraint is satisfied
depends on the value of $s_0$.  The comparison between pole and
continuum contributions for $s_0^{1/2} = 4.2$ GeV is shown in
Fig.~\ref{figpvc}.
The same analysis for the other value of the continuum  
threshold gives $M^2 < 2.2$  GeV$^2$ for $s_0^{1/2} = 4.1~\GeV$.

In Fig.~\ref{figmx}, we show the $X$ meson mass obtained from
Eq.~(\ref{m2}), in the relevant sum rules window, with the upper and
lower validity limits indicated.  From Fig.~\ref{figmx} we see that
the results are reasonably stable as a function of $M^2$.
In our numerical analysis, we shall then consider the range of $M^2$ values
from 2.0 $\GeV^2$ until the one allowed by the sum rule window criteria as 
can be deduced from Fig.~\ref{figmx} for each value the $s_0-$ range of values. 

Using the QCD parameters in Eq. (\ref{qcdparam}), we obtain the LSR 
predictions for different values of $s_0$ and including the dimension-5 
condensates:
\beqa
M_X &=
&(3908\pm 26 \pm 13 \pm 100 \pm 46) \MeV~~~{\rm for} ~~~s_0^{1/2}= 4.1~
{\rm GeV},
\nnb\\
&& (3943\pm 30 \pm 10 \pm 80 \pm 48) \MeV~~~{\rm for} ~~~s_0^{1/2}= 4.2~
{\rm GeV}.
\enqa

The errors are due  respectively to $M^2,~\langle\bar{q}q\rangle$,
$m_c$ and  the truncation of the OPE. We have estimated the absolute value 
of the last error by varying the 
dimension-six and eight condensates  from their vacuum saturation values to 
the ones where a violation of the factorization
assumption by a factor two is assumed.  The errors  due to other parameters 
are negligible. One can notice 
that the main error comes from the uncertainties in the determination of the 
charm quark mass, which plays a crucial role in the analysis like in the one 
of other heavy quark systems. One can also notice that the central value of 
the mass prediction increases with $s_0$. Apart the intuitive observation 
from an extrapolation of the known mass splittings from ordinary mesons 
which may not be applied for the multi-quark states (see e.g.\cite{MATTHEUS}),
$s_0$ remains a free parameter. We shall  try to fix its value using FESR.

\boldmath \section{FESR prediction for $M_X$} \unboldmath
As an alternative, we use the FESR, which can be obtained from
Eq.~(\ref{sr}) by taking the limit ${1}/{M^2}\rightarrow 0$ and equating the 
same power in $1/M^2$ in the
two sides of the sum rules.  We get up to dimension-six condensates:
\beq
2f_X^2M_X^8 \sum_{n} (-M_X^2)^{n} \left(\frac{1}{M^2}\right)^{n} 
=
\sum_{n} \int_{4 m_c^2}^{s_0} ds  (-s)^{n}
\left(\frac{1}{M^2}\right)^{n}  
\rho(s), \;\;\;  {n= 0,1,2...}.
\label{fens}
\enq
Equating the
coefficients of the polynomial in $1/M^2$ in both sides of
Eq.~(\ref{fens}) gives $n$ equations:
\beq
2f_X^2M_X^8 M_X^{2n}  
=
\int_{4 m_c^2}^{s_0} ds \; s^{n} \rho(s), \;\;\;  n=0,1,2...
\enq
Finally, dividing two subsequent equations (with $n$ and $n+1$), we can
obtain the mass $M_X$ for any chosen value of $n$ (which, formally, is 
expected to by the same for any $n$):
\beq
M_X^2=\frac{\int_{4 m_c^2}^{s_0} ds \; s^{n+1} \rho(s)}{\int_{4
m_c^2}^{s_0} 
ds \; s^{n} \rho(s)}, \;\;\;  n=0,1,2...
\label{fesr.n}
\enq

\begin{figure}[h] 
\centerline{\epsfig{figure=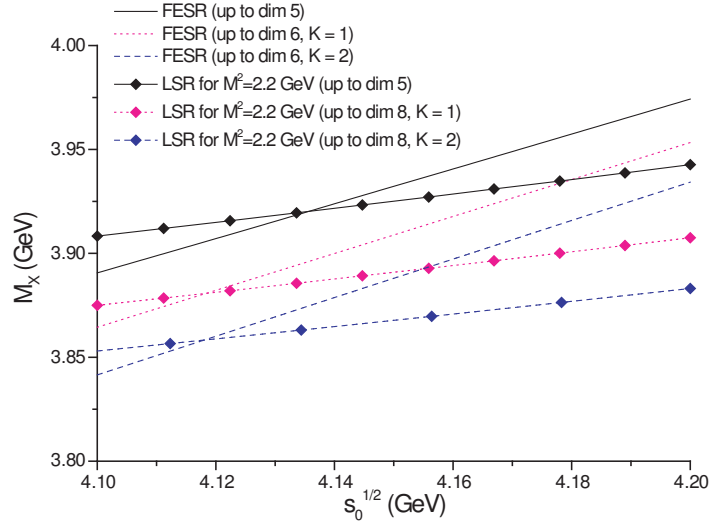,height=70mm}}
\caption{The FESR results in Eq.~(\ref{fesr.n}) for $M_X$ as a
function of $s_0$ for $n=0$ including condensates up to different dimensions.
The LSR results have been also inserted for easy comparison. }
\label{fesr.eps} 
\end{figure}

In contrast to the previous method, the FESR have the advantage of
giving correlations between the mass and the continuum threshold
$s_0$, which can be used to avoid inconsistencies in the determination
of these parameters.  Ideally, one looks at a minimum in the function
$M_X(s_0)$, which would provide a good criteria for fixing both $s_0$
and $M_X$.  The results for different values of $n$ are very similar, 
therefore, in Fig.~\ref{fesr.eps}, we only show the result for $n=0$. One
can see in Fig.~\ref{fesr.eps} that there is no stability in $s_0$, which
can indicate the important role of the QCD continuum in the analysis.

One can also notice, from Fig.~\ref{fesr.eps},
that the FESR converges faster than the LSR due mainly to 
the fact that here we use an expansion in $1/s_0$ where $\sqrt{s_0}\sim 4.1$ 
GeV, while in the LSR, the expansion is done in $1/M^2$, where $M\sim1.4$ GeV 
is much smaller.
{\boldmath \section{Final QSSR predictions for $M_X$ and $f_X$} \unboldmath
In order to exploit the complementary role of the LSR and FESR, we
also show in Fig.~\ref{fesr.eps} , the result
obtained from the Laplace sum rule using $M^2=2.2~\GeV^2$. The factor $K$ was
introduced to account for deviations of the factorization hypothesis 
 for the $D=6$ \cite{FACTO,DOSCH,RAF,GIMENEZ,SNG,ALEPH} and $8$ 
\cite{BAGAN,RAF,SNG,ALEPH} 
$-$condensates; $K=1(2)$ refers to an assumption that respects (violates by 
a factor 2) vacuum saturation.

The intersection point fixes the range of values of $s_0$ to be:
\beq
s_0^{1/2}=(4.15\pm 0.03)~{\rm GeV}~,
\enq
which is smaller than intuitively expected. This small value of 
the continuum
threshold relative to the value of the resonance mass signals again the 
important role of the QCD continuum in the analysis, which is expected
for correlators described by high-dimension current operators.  Using this 
range of values of $s_0$, one can, definitely, fix  the $X$ mass to be:
\beq\label{eq: finalX}
M_X= ( 3925\pm 20 \pm 46\pm 117 )~{\rm MeV}~,
\enq
in remarkable agreement (within the errors) with the experimental candidate 
$X(3872)$ and with an estimate from relativistic quark model 
\cite{EBERT}\footnote{However, due to the
large error of our result, it can also be compatible with the $X(3940)$,  
$Y(3940)$ and  $Z(3930)$ if some of them are found to be a $1^{++}$ state. We 
plan to study carefully the splitting
of the different states of a given spin in a future work.}. 
The first error comes from $s_0$, the second from the truncation of the OPE,
 and the third from the QCD inputs, such as $m_c$ and $\lag\bar{q}q\rag$. 
Despite this large 
dependence of our results on the value of the continuum threshold, 
the error induced by $s_0$ is comparable with the ones from 
other sources. The errors
due to $s_0$ could be reduced by using a more involved parametrization based on some
 effective Lagrangian
or eventually, alternatives forms of the sum rules, such as those used in  
\cite{BNP}  for describing the hadronic $\tau$ decay or some other sum rules \cite{MALT2}  .

Assuming that the mass of the first radial excitation is given by 
$\sqrt{s_0}$, one can deduce a crude estimate of the splitting:
\beq
X'-X\approx 225 ~{\rm MeV}~,
\label{x'x}
\enq
which is expected to be valid if the local quark-hadron duality is 
at work\footnote{If one uses similar assumption  for the $D_s^J(0^{++})$, 
one can identify $\sqrt{s_0}$ with the radial excitation predicted in 
\cite{SWANSON2} using some other approaches.}.
 Within this assumption, one can
notice that the mass-splitting is much smaller that the na\" \i ve 
extrapolation from the ordinary meson 
spectrum. Such a situation has been also
encountered in the analysis of the pentaquark sum rule \cite{MATTHEUS}
and, in general, in the analysis of correlators described by
high-dimension operators such as hybrids and gluonia \cite{SNB}. This result
can indicate the existence of  higher states near the lowest ground state 
mass, which can manifest as large continuum in the data analysis.

One can also deduce to leading order in $\alpha_s$, from the individual 
lowest moments, the decay constant defined in Eq.~(\ref{eq: decay}):
\beq\label{fx}
f_X=(4.66\pm 0.16\pm0.29\pm0.68)\times 10^{-5}~{\rm  GeV}~,
\enq
which can be more affected by the radiative corrections than $M_X$. 
The first error comes from $s_0$, the second from the truncation of the OPE,
and the third from the QCD inputs, such as $m_c$ and $\lag\bar{q}q\rag$. 
$f_X$ is 
useful for the estimate of its hadronic width using vertex sum rules. 
As $X$ is an axial-vector meson, its decay constant can measure its weak 
transition into $l\nu$ via a $W$-exchange,which might be difficult to measure 
experimentally.
It would be useful to have a measurement of this decay constant from some 
other methods, like e.g. lattice calculations.
\boldmath \section{SU(3) breakings and mass of the $X^{s}$}\unboldmath
%
\begin{figure}[h] 
\centerline{\epsfig{figure=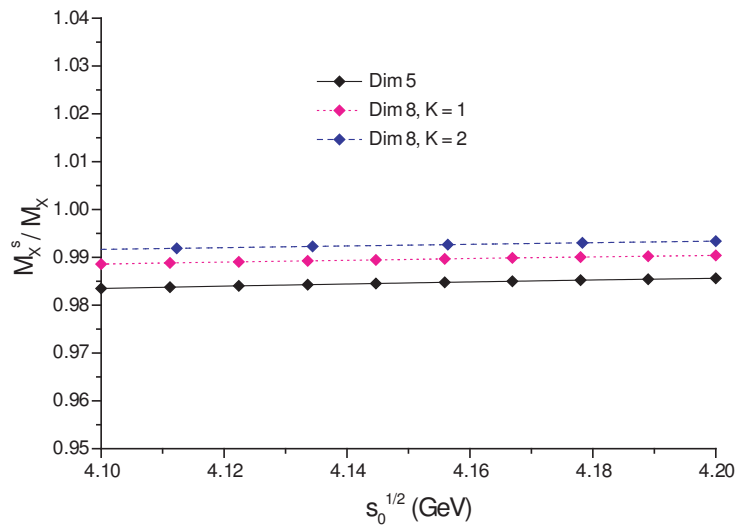,height=70mm}}
\caption{The ratio $M_{X^s}/M_X$ as a function of $s_0$, obtained from the 
LSR results up to different dimensions in the OPE.}
\label{fesrs.eps} 
\end{figure}

It is straightforward to extend the previous analysis to the case of the 
strange quark by using the QCD parameters given in Eq. (\ref{qcdparam}). 
One can e.g. work with the ratios given in Eqs. (\ref{m2}) and (\ref{fesr.n}).
However, as the errors in the determination of $M_X$ are relatively large,
it will be difficult, to extract the SU(3) splitting from these individual 
ratio of moments.

For extracting this relatively small mass-splitting, it is appropriate 
to use the double
ratio of moments \cite{SNBB,SNB}:
\beq\label{doubleratio}
{d}_{c}^s\equiv  {M^2_{X^s}\over M^2_X}
\enq
for the LSR and for the FESR, which suppress different systematic errors 
($m_c,$...) and 
the dependence on the sum rule parameters ($s_0, M^2$). The results of the 
analysis from LSR are given in Fig. \ref{fesrs.eps}
from which one can deduce, with a good accuracy:
\beq
\sqrt{d^s_c}= 0.984\pm 0.002 \pm 0.007~,
\enq
where the first error comes from the QCD and sum rules parameters  including the $SU(3)$ breaking
of the quark condensates. The 
second error
from an estimate of the truncation of the OPE. This leads to the mass 
splitting:
\beq
M_{X^s}-M_X\simeq-(61\pm 30)~\MeV~.
\label{mxs}
\enq
Similar methods used in \cite{SNBB,SNB} have predicted successfully the
values of $M_{D_s}/M_D$ and $M_{B_s}/M_B$, which is not quite surprising,
as in the double ratios, all irrelevant sum rules systematics cancel out. 

Using FESR and taking the SU(3) breaking correction for the continuum 
threshold, which is the most important effect in the FESR analysis, we confirm the previous 
LSR result.
It is interesting to notice that we predict a $X^s$ mass slightly lighter 
than the $X$,
which is quite unusual. This is due to the fact that, in the sum rules expression 
of $M^2_{X^s}$, 
the linear quark mass term 
tends to decrease the $X^s$-mass, which is partly compensated
by the effect of the quark condensates. Such a small and negative 
mass-splitting is rather striking
and needs to be checked using alternative methods. Note,
however, that a partial restoration of SU(3) symmetry  is already
observed in the neighbourhood of heavy quarks, illustrated by the almost 
equal hyperfine splittings $D_s^*-D_s$ and $D^*-D$. In potential models,
the mass spliting $M_{X^s}-M_X$ is certainly larger than the value in
Eq.~(\ref{mxs}), but smaller than $2(m_s-m_q)$ as the increase of the 
constituent mass from $m_q$ to $m_s$ is partially cancelled by  the
deeper binding of the strange quarks. The existence of the $X^s$, which can 
be experimentally checked, can serve for a further test of the four-quark 
model for the $X$. The (almost) degenerate value of the $X$ and of the $X^s$
masses may suggest that the physically observed $X$ state can result from a 
mixing
between the $c\bar c q\bar q$ and $c\bar c s\bar s$ bare states, which may
be dominated by its $c\bar c q\bar q$ component. However, we expect that
a careful and perfect analysis of the  $c\bar c s\bar s$ sector should feel
the $X$ in the spectrum, though with a small coupling. One should also notice 
that  these $c\bar c q\bar q$ and $c\bar c s\bar s$ components can be
comparable if the $X$ is a $SU(3)$ singlet state.

Using  the ratio of the $s$- over the $q$-quark sum rules , one can predict 
also the ratio of decay constants:
\beq\label{ratiofxsfx}
{f_{X^s}\over f_{X}} \simeq  1.025\pm0.010 
\enq
where, in the individual sum rules, the $m_s$ corrections act positively 
implying that this ratio is larger than 1.  We also expect the reliability of
a such result advocating the previous arguments for the ratio of mass. 
Similar sum rule leads to
${f_{B^s}/ f_{B}}=1.16\pm 0.03$ \cite{SNFB}, which has been confirmed later on 
by different lattice calculations.

However, despite the different successful predictions of the ratio of moments 
for the $B$-meson parameters, we expect that the method will be less predictive 
for the four-quark state. This can be signaled by the large
error in the previous prediction of the mass-difference. The inclusion 
of radiative or some other higher dimension
condensates corrections or some other effects not accounted for in 
this paper will be useful for confirming or
disproving the previous results.
\section{Test of the isospin violation}
We attempt to use of the sum rule, for  a rough
estimate of the small mass difference $M(X_h) - M(X_l)$ defined in 
Eq.~(\ref{difma}).  Using
Eq.~(\ref{m2}), we get: 
\beq M^2(X_h) - M^2(X_l) = \frac{\int_{4
m_c^2}^{s_0}ds ~e^{-s/M^2}~s~
\left[\rho_h(s)-\rho_l(s)\right]}{\int_{4 m_c^2}^{s_0}ds
~e^{-s/M^2}~\rho(s)},
\label{m2mm2}
\enq
where:
\beq
\rho_l(s)=\cos^2\theta ~\rho_u(s) + \sin^2\theta ~\rho_d(s)\quad
\mbox{and}\quad
\rho_h(s)=\sin^2\theta ~\rho_u(s) + \cos^2\theta  ~\rho_d(s).
\enq
Here, $\rho_u(s)$ and $\rho_d(s)$ are simply the spectral density
$\rho(s)$ defined before with the flavor of the light quark chosen as
$u$ and $d$, respectively.

Clearly the only terms depending on the light quark flavor
will contribute to the numerator of Eq.~(\ref{m2mm2}).  In fact the
expression $\rho_h(s)-\rho_l(s)$ can be written in terms of the
isospin breaking quantities: $\uu - \dd = - \gamma \qq$, $m_u - m_d$,
$m_u \uu - m_d \dd = - \gamma \qq m_q + \qq (m_u - m_d)$, and
$\uu^2-\dd^2= -2 \gamma \qq^2$, where $\qq={(\uu + \dd)/ 2}$ and
$\ga={(\lag0|\bar{d}d-\bar{u}u|0\rag)\lag0|\bar{u}u|0\rag}$.

The value of $\ga$ has been estimated in a variety of approaches 
with results varying over almost one order of 
magnitude: $-1\times10^{-2}\leq\ga\leq-2\times10^{-3}$ \cite{jim}. However,
studies based on 
chiral perturbation theory \cite{gale}, LSR  \cite{SNB} and FESR \cite{DERAF} analysis of the (pseudo)scalar 
channels , 
analysis of the neutron-proton mass difference \cite{jim} and of the heavy
meson decay widths \cite{dsdpi} leads to the value:
\beq
\ga\simeq -(1\pm 0.5) \times 10^{-2}
\label{gamma_a}
\enq
which we shall consider in our analysis .  The results for the
mass difference using $s_0^{1/2} = 4.2$ GeV can be seen in
Fig.~\ref{figuvsd}, where we have considered two values for the mixing
angle: $\theta = 0^\circ$, corresponding to  maximal isospin violations, 
and
$\theta = 20^{o}$ which was the value determined in \cite{Maiani}.

\begin{figure}[h] 
\centerline{\epsfig{figure=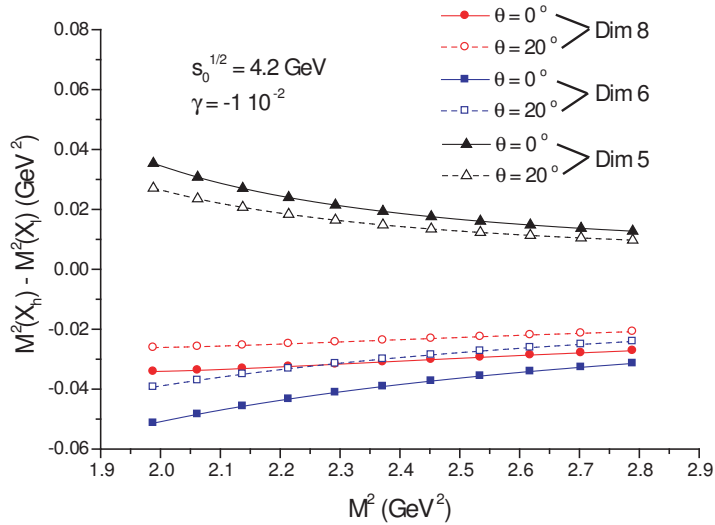,height=70mm}}
\caption{$M^2(X_h) - M^2(X_l)$ as a function of the sum rule parameter
($M^2$) for $\gamma = -1\times10^{-2}$.  The solid line is for $\theta
= 0^{o}$ and the dashed one is for $\theta = 20^{o}$.}
\label{figuvsd} 
\end{figure}

On can notice from Fig.~\ref{figuvsd} that the sign of the mass 
difference is reversed when one includes
the dimension-six condensates, while the effect of the (partial) dimension-8 
contribution is relatively small,
indicating that the OPE starts to behave quite well. However, one needs a more 
complete  evaluation of the
dimension-8 contributions for a more precise determination of the 
mass-spiltting. For a more conservative estimate, we consider the range of the 
absolute value of the mass-difference, which is not strongly affected by the 
truncation of the OPE. In this way, one obtains  from  Fig.~\ref{figuvsd}: 
\beq\label{eq: split2}
|M(X_h)-M(X_l)|\simeq ( 2.6~-~3.9){\rm MeV}~,
\enq
which is smaller than the  $(8\pm 3)$ MeV value given in \cite{Maiani}, but larger than 
the decay width of the $X(3872)$, which is less than 2.3 MeV. However, one can 
notice from  Fig.~\ref{figuvsd} that the sum rule cannot fix with a good 
precision the sign of the mass splitting, though it is
tempting to conclude that the sign is negative, in disagreement with the result 
of \cite{Maiani}.  
\boldmath \section{Sum rule predictions for $X_b$ and $X^s_b$} \unboldmath
Using the same interpolating field of
Eq.~(\ref{field}) with the charm quark replaced by the bottom one, the 
analysis done for $X(3872)$ in the previous sections can be
repeated for $X_b$, where $X_b$ stands for a $(b\bar{b}q\bar{q})$
tetraquark axial meson.
Using consistently the perturbative $\overline{MS}$-mass $m_b(m_b)=4.24~\GeV$, and working 
with the LSR, we find a good OPE convergence for
$M^2>5~\GeV^2$.  We also find that, for $s_0<(10.2~\GeV)^2$, the
continuum contribution is always bigger than the pole contribution for
all values of $M^2>5~\GeV^2$.

\begin{figure}[h] 
\centerline{\epsfig{figure=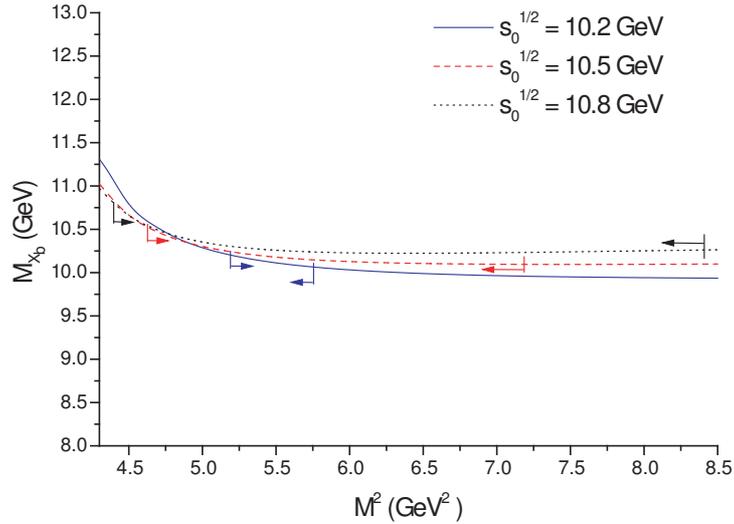,height=70mm}}
\caption{The $X_b$ meson mass as a function of the sum rule parameter
($M^2$) for different values of the continuum threshold:
$\sqrt{s_0}=10.2$ GeV (solid line), $\sqrt{s_0}=10.5$ GeV (dashed
line) and $\sqrt{s_0}=10.8$ GeV (dotted line).  The arrows delimit
the region allowed for the LSR sum rules.}
\label{figmxb} 
\end{figure} 

In Fig.~\ref{figmxb} we show the $X_b$ meson mass obtained from
Eq.~(\ref{m2}), in the relevant sum rules window, with the upper and
lower validity limits indicated.  Although we get a good OPE
convergence for $M^2>5~\GeV^2$, we have now a more restricted lower
limit given by $M_{X_b}<\sqrt{s_0}$.  Therefore, the lower limit
indicated in Fig.~\ref{figmxb} is given by this condition.

From Fig.~\ref{figmxb} we see that the results are very stable as a
function of $M^2$ in the allowed region.  However, the LSR prediction
increases with $s_0$. Taking into account the
variation of $M^2$ and choosing (a priori) some range of $s_0$, we arrive at 
the predictions:
\beq
\lb{massXb}
10.06~\GeV \leq M_{X_b}\leq   10.50~\GeV~,
\enq
for  $10.2~\GeV \leq {s_0}^{1/2}\leq10.8$ GeV and $5.0\leq M^2\leq 8.5~\GeV^2$.
\begin{figure}[h] 
\centerline{\epsfig{figure=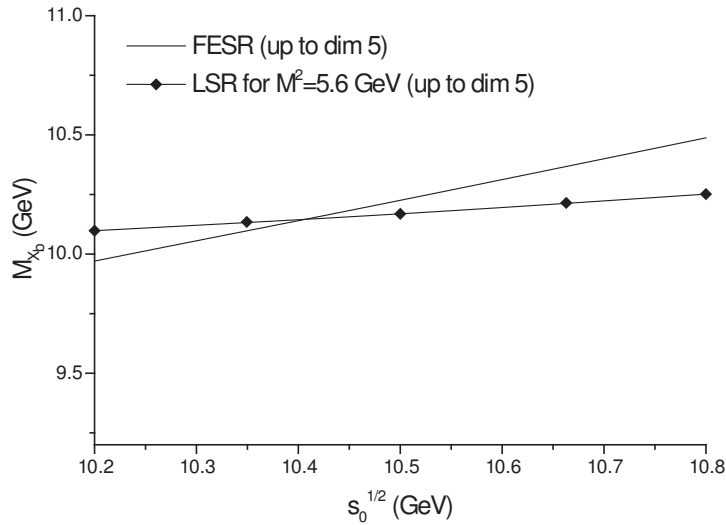,height=70mm}}
\caption{The FESR results in Eq.~(\ref{fesr.n}) for $m_{X_b}$ as a
function of ${s_0}^{1/2}$ for $n=0$.  The LSR result
with $M^2 = 5.6 \GeV^2$ has been also inserted for easy comparison.}
\label{cross.eps} 
\end{figure}

The FESR analysis can also be repeated in the case of 
the $b$-quarks for improving the LSR results. The results are shown in 
Fig.~\ref{cross.eps}. As 
in the case of $X(3872)$ the curves for $n=0$ and $n=1$ are quite
similar, and again there is no stability in $s_0$. 
We also show in the same figure the LSR results for $M^2 = 5.6 \GeV^2$, as
a function of ${s_0}^{1/2}$. A common solution is obtained for:
\beq
{s_0}^{1/2}=(10400\pm 20)~{\rm MeV}~,
\enq
to which corresponds the improved final prediction:
\beq\label{eq: finalX_b}
M_{X_b}= (10144\pm 21  \pm 104)~{\rm MeV}~.
\enq
The second error comes again from the QCD inputs.
The central value in Eq.~(\ref{massXb}) is close to the mass of
$\Upsilon(3S)$, and appreciably below the $B^*\bar{B}$ threshold at
about $10.6\;$GeV. For comparison, the molecular model predicts for
$X_b$ a mass which is about $50-60\;$MeV below this threshold
\cite{Swanson}, while a relativistic quark model without explicit 
$(b\bar{b})$ clustering predicts a value 
of about 133 MeV below this threshold \cite{EBERT}. 
It would also be 
interesting to have the (unquenched) lattice  results for this state in 
order 
to test our QCD based results. A future
discovery of this state, e.g. at LHCb, will certainly test the different 
theoretical models on this state and clarify, in  the same time, the 
nature of 
the $X(3872)$.  

One can also notice, by assuming that the mass of first radial excitation is 
about the value of $\sqrt{s_0}$:
\beq
X'-X\approx X'_b-X_b \approx (225\sim250)~{\rm MeV}~.
\enq
For completeness, we predict the corresponding useful value of the decay 
constant to leading order in $\alpha_s$:
\beq\label{fxb}
f_{X_b}= (6.9\sim 7.1)\times 10^{-6} ~{\rm GeV}~.
\enq
Our previous results will be useful inputs for studying more precisely the 
phenomenology of the $X_b$ outlined in \cite{WHOU}.
\begin{figure}[h] 
\centerline{\epsfig{figure=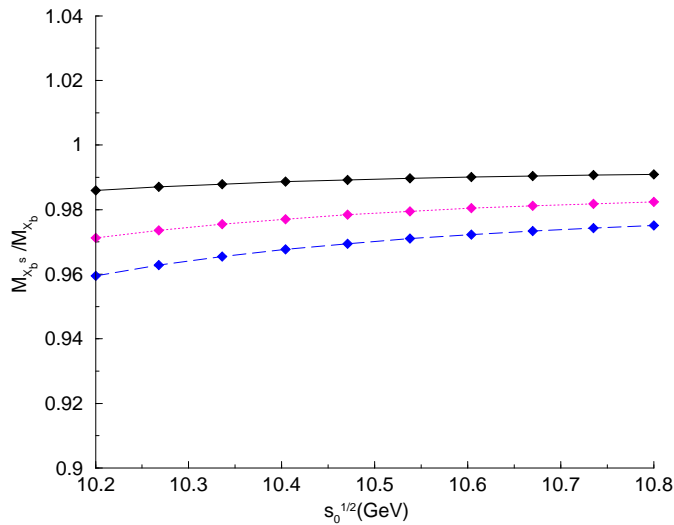,height=70mm}}
\caption{The ratio $M_{X^s_b}/M_{X_b}$ as a function of $s_0$, obtained from 
the LSR results up to different dimensions in the OPE. The continuous curve 
corresponds to the $D=5$ condensate
contributions. The dotted and dashed curves correspond respectively to the 
vacuum saturation ($K=1$) and violation of the vacuum saturation ($K=2$) by a 
factor 2 of the dimension six- and (partial) eight-contributions.}
\label{ratiob.eps} 
\end{figure}

We extend the analysis to the $X_b^s(b\bar{b}s\bar{s})$. We show in 
Fig.~\ref{ratiob.eps}  the LSR prediction for the
mass ratio, from which we deduce, by truncating the OPE at $D=5$:
\beq
\sqrt{d_b^S}\equiv  {M_{X^s_b}\over M_{X_b}}=0.988\pm 0.002\pm 0.018~,
\enq
where the first error comes from $s_0$ and the QCD parameters, while the second one from the 
truncation of the OPE.
This leads to:
\beq
{M_{X^s_b}- M_{X_b}} = -(123\pm 182) ~{\rm MeV}~,\qquad{\rm and}\qquad
{f_{X^s_b}\over f_{X_b}}\simeq 1.12\pm 0.03~,
\label{eq:split3b}
\enq
where the error due the truncation of the OPE is larger than in the case of the 
$c$-quark, which is mainly due
to the terms of the form $m_b^2/M^2$ in the OPE.  We expect that the $X_b$-family
will show up at LHCb in the near future, which will serve as a test of our 
previous predictions. 
\section{Conclusions}
We have presented a QSSR analysis of the two-point
functions of the $X(3872)$ meson considered as a four quark state.  We
find that the sum rules result in Eq.~(\ref{eq: finalX}) is compatible 
with experimental data. An improvement of this result needs an accurate
determination of running mass $m_c$ of the ${\overline{MS}}$-scheme and the inclusion of radiative
corrections. 

We have extended the analysis for studying the mass splitting between the
$X^s$ and $X$ due to SU(3) breaking. Our result in Eq.~(\ref{mxs}) indicates
an unusual ordering which deserves further independent checks from
other QCD-based approaches, especially lattice calculations.  However,
our small mass-splitting  suggests that perhaps the
observed $X$ has a $c\bar c s\bar s$ component, though with a small
coupling, with a size which depends on the SU(3) assignment of the $X$.

Allowing possible isospin violations, we have also studied the mass
splitting between the states $X_l$ and $X_h$. Using the common values 
from different approaches of the 
leading SU(2) breaking parameter $\gamma$ of the light quark 
condensates defined in Eq. (\ref{gamma_a}),  we obtain  the splitting 
in Eq. ~(\ref{eq: split2}) which is smaller than the value
$8~\mbox{MeV}$ predicted in \cite{Maiani}.

There are limits \cite{Aubert:2004zr} on the production of charged partners 
of the $X(3872)$, but based on the weak decay of $B$ mesons. It cannot be 
excluded that $B$ decay favours neutral heavy mesons, if it proceeds first 
via an excited $(c\bar{c})$ state recoiling against a cluster of light quarks
 and antiquarks. Hence the search for charged partners should be extended 
to other production mechanisms.

Extending our analysis to the $b$-quark meson, we found the values of the  
$X_b$ and $X_b^s$ masses in Eqs. (\ref{eq: finalX_b}) and 
(\ref{eq:split3b}), which are  appreciably below the  
$B^*\bar{B}$ threshold at about $10.6\;$GeV.  This is a common feature of 
all quark models with proper account for the correlation between the heavy 
quark and the heavy antiquark that the $X_b$ is more deeply bound with 
respect to  $B\bar{B}{}^*$ than $X_c$ with respect to $D\bar{D}{}^*$, for 
the same reason why the $(b\bar{b})$ family has more narrow states than 
$(c\bar{c})$. In contrast, the molecular model, in which $X_b$ is a 
meson--meson system bound by nuclear forces predicts this state rather 
close below the $B\bar{B}{}^*$ threshold.
Our analysis also indicates that the mass-splitting between the ground 
state and the first radial
excitation is about ($225\sim 250$) MeV, which is much smaller than the one 
expected from ordinary mesons,
and which are (within the errors) flavour independent.

We present in Eqs. (\ref{fx}), 
and (\ref{fxb}) 
predictions of the decay constants  of the $X$ and $X_b$,
and in Eqs. (\ref{ratiofxsfx}) and (\ref{eq:split3b}) the ratio of the 
strange 
over the non strange decay constants. These are useful quantities for
studying the leptonic and hadronic decay widths of such mesons, and which
can be checked from (unquenched) lattice calculations or from some other
models.

A future discovery at B factories or LHCb of the different states which we 
have predicted as a consequence of the
$1^{++}$ four-quark nature assumption of the $X(3872)$  will certainly test 
the different theoretical models proposed for this state and clarify, in 
the same time, the nature of the $X(3872)$. 

Different choices of the four-quark operators have been systematically
presented for the $0^{++}$ light mesons in  \cite{chinois}, which should mix 
under renormalizations \cite{TARRACH}
from which one can deduce a ``physical" renormalization group invariant 
current (RGI) which 
can describe the observed state. Though some combinations can 
provide a faster convergence of the OPE, we do not 
expect that the choice of the operators
will affect much our results, where, in our analysis, the OPE has a good 
convergence while the 
renormalization mixing is a higher order effect in $\alpha_s$.
Another choice of operator not included in the previous analysis is, for
instance, given in ref.~\cite{hrs}, where it was shown that a simple 
chromomagnetic
model suggests that the color octet-octet in the $[c\bar{c}]_{S=1}
[q\bar{q}]_{S=1}$ basis is the most natural candidate for describing
the $X(3872)$.  However, this choice would correspond to an operator
of higher dimension than the one analyzed in Eq.~(\ref{field}), which
would therefore induce relatively small corrections to the present
analysis. Though done with a particular choice of current \cite{Maiani},
we expect that the results given in this paper will reproduce (within the 
errors of the approach) the general features
of the four-quark model for the $X(3872)$.  We plan to come back to these
issues in a future publication.

Once the mass of the $X(3872)$ is understood, it remains to explain
why it is so narrow.  There are presumably many multiquark states, but
most of them are very broad and cannot be singled out from the
continuum.  In a recent study \cite{nani}, based on the same
interpolating field as the one used here, it was shown that, in order to 
explain
the small width of the $X(3872)$, one has to choose a particular set
of diagrams contributing to its decay.  However, it will be desirable if 
this investigation can be
checked from alternative approaches, such as lattice  
calculations. If confirmed, this method can be straightforwardly  
repeated to a variety of currents for understanding the
width and the internal structure  of the $X(3872)$.

If the $X(3872)$ is  a four-quark state, as our analysis 
suggests in answer to the question raised in the title, a four-quark 
structure probably holds for the states seen near $3940\;$MeV and 
$4260\;$MeV, on which more experimental information is still needed. This 
is our intention to extend the present analysis to other $J^{PC}$ configurations
 which are likely to host multiquark resonances.

\section*{Acknowledgements}
{This work has 
been partly supported by the CNRS-IN2P3 within 
the (Non) Exotic Hadrons Working Group Program, and by FAPESP and 
CNPq-Brazil.} S.N. wishes to thank the CERN-Theory Division where this work has been completed.

\end{document}